\documentclass[journal=jctcce,manuscript=article]{achemso}
\setkeys{acs}{articletitle = true}
\usepackage{chemformula}
\usepackage[T1]{fontenc}
\usepackage{blindtext}

\newcommand{\onlinecite}[1]{\hspace{-1 ex} \nocite{#1}\citenum{#1}}
\def\half{ \frac{1}{2}}

\author{Micha\l\ Lesiuk}
\email{lesiuk@tiger.chem.uw.edu.pl}
\affiliation{\sl Faculty of Chemistry, University of Warsaw, Pasteura 1, 02-093 Warsaw, Poland}
\date{\today}

\title[]{Implementation of the full CCSDT electronic structure model with tensor 
decompositions}


\begin{document}

\begin{tocentry}
\includegraphics[scale=0.66]{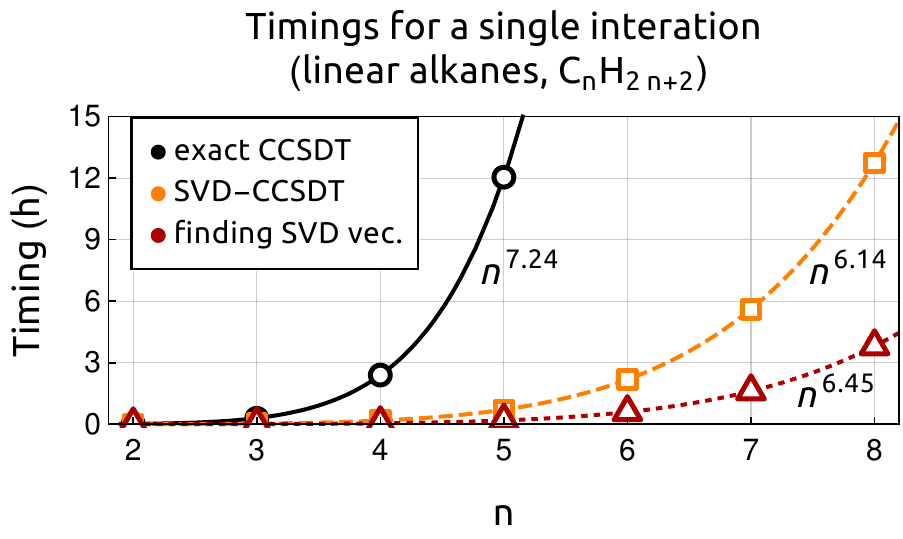}
\end{tocentry}

\renewcommand{\arraystretch}{1.5}

\begin{abstract}
We report a complete implementation of the coupled-cluster method with single, double, and triple excitations (CCSDT) 
where tensor decompositions are used to reduce its scaling and overall computational costs. For the decomposition of 
the electron repulsion
integrals the standard density fitting (or Cholesky decomposition) format is used. The coupled-cluster single and 
double amplitudes are treated conventionally, and for the triple amplitudes tensor we employ the Tucker-3 compression 
formula, $t_{ijk}^{abc} \approx t_{XYZ} \,U^X_{ai}\,U^Y_{bj} \,U^Z_{ck}$. The auxiliary quantities $U^X_{ai}$ come from 
singular value decomposition (SVD) of an approximate triple amplitudes tensor based on perturbation 
theory. The efficiency of the proposed method relies on an observation that the dimension of the ``compressed'' 
tensor $t_{XYZ}$ sufficient to deliver a constant relative accuracy of the correlation energy grows only 
linearly with the size of the system, $N$. This fact, combined with proper factorization of the coupled-cluster 
equations, 
leads to practically $N^6$ scaling of the computational costs of the proposed method, as illustrated 
numerically for linear alkanes with increasing chain length. This constitutes a considerable improvement over  
the $N^8$ scaling of the conventional (uncompressed) CCSDT theory. The accuracy of the proposed method is verified
by benchmark calculations of total and relative energies for several small molecular systems and comparison 
with the exact CCSDT method. The accuracy levels of 1 kJ/mol are easily achievable with reasonable SVD subspace size, 
and even more demanding levels of accuracy can be reached with a considerable reduction of the computational costs. 
Extensions of the proposed method to include higher excitations are briefly discussed, along with possible strategies 
of reducing other residual errors.
\end{abstract}

\newpage
\section{Introduction}
\label{sec:intro}

Due to numerous favorable properties such as the rigorous size-extensivity, polynomial scaling of the 
computational costs with 
the system size and rapid convergence toward the full configuration-interaction (FCI) limit, the coupled cluster (CC) 
theory \cite{coester58,coester60,cizek66a,cizek66b,cizek71,cizek72} has become one of the most important theoretical 
tools for 
finding approximate solutions of the electronic Schr\"{o}dinger equation for many-electron atoms and molecules (see 
Refs.~\onlinecite{crawford00,bartlett07} for exhaustive recent reviews). In particular, a variant of the CC theory with 
single and double 
excitations included in the wavefunction \cite{purvis82,scuseria87} and a perturbative treatment of the triple 
excitations \cite{ragha89}, named CCSD(T), offers 
a very advantageous accuracy-to-cost ratio and serves as the \emph{gold standard} of computational chemistry. 
However, in applications where, e.g., accuracy better than 1 kcal/mol is 
required\cite{csaszar98,halkier99,sordo01,harding07,vazquez09,feller09} or the 
wavefunction is no 
longer dominated by a single reference determinant 
\cite{chan58,laidig87,kowalski00,piecuch02,piecuch04,shen12}, the CCSD(T) model 
becomes inadequate. In such 
cases it is reasonable 
to climb further up the coupled-cluster ladder and consider the full CCSDT method \cite{noga87,scuseria88} with 
optional perturbative 
corrections, akin to (T), to take higher excitations into account 
\cite{kucharski89,kucharski98,kucharski01,bomble05,kallay05,kallay08}. Unfortunately, there is a huge gap in the 
computational costs between the two rungs of the ladder, i.e., with $N$ being the size of the system, the number of 
operations needed to calculate the most expensive term in the CCSD(T) equations scales as $N^7$ while in 
the CCSDT equations -- as $N^8$. In practical terms, this translates into orders of magnitude difference in 
computational timings for realistic systems.

The reasons for this substantial increase of the computational costs are directly linked to the 
presence of the coupled-cluster triple excitation amplitudes tensor, $t_{ijk}^{abc}$, which can no 
longer be neglected or treated implicitly (details of the notation are given further in the text). The quantity 
$t_{ijk}^{abc}$ is treated here as a fully symmetric rank-3 
tensor with compound indices $ai$, $bj$, and $ck$. If one denotes the number of correlated occupied and 
virtual orbitals by $O$ and $V$, respectively, the dimension of this tensor is $OV$. Despite the number 
of unique elements of the $t_{ijk}^{abc}$ tensor is very large, roughly $\frac{1}{6}O^3V^3$, one 
can assume that only a comparatively small number of excitations (or linear combinations thereof) 
is actually important for the quality of the results. It is thus reasonable to employ 
rank-reduction or ``compression'' techniques that decompose the full tensor into quantities of 
a lower rank, thereby reducing the computational burden and storage requirements of the method. In 
this paper we rely on the Tucker-3 decomposition\cite{tucker66} which represents the triple excitation amplitudes 
tensor in the following form
\begin{align}
\label{tuck1}
&t_{ijk}^{abc} \approx t_{XYZ} \,U^X_{ai}\,U^Y_{bj} \,U^Z_{ck},
\end{align}
where the summation over repeating indices is implied, as is throughout the present work.
This compression method is successful when the dimension of the tensor $t_{XYZ}$, denoted $N_{\mathrm{SVD}}$ for 
reasons explained further in the text, is 
significantly smaller than the initial one, $OV$.  The 
first application of the above formula in the coupled-cluster calculations was presented by Hino, 
Kinoshita and Bartlett\cite{hino04} at the $\mbox{CCSDT-1}$\cite{lee84,noga87b} level of theory (an approximate 
variant of CCSDT). Despite the results 
were very promising, a 
$N^8$ process was necessary to obtain the $U^X_{ai}$ tensors, i.e., the method was 
formally as expensive as the full CCSDT calculations. However, this obstacle has recently been  
removed\cite{lesiuk19} and a suitable set of $U^X_{ai}$ tensors can now be obtained much more cheaply by iterative 
singular-value decomposition of some approximate $t_{ijk}^{abc}$ amplitudes. This enables us to apply 
the decomposition (\ref{tuck1}) at 
the CCSDT level of theory which is the main purpose of this work. Let us also stress that the 
present paper reports the first application of the decomposition (\ref{tuck1}) in a coupled-cluster 
method where the triply excited amplitudes must be explicitly formed and stored (in contrast to the 
CCSDT-1/CC3 methods where the $t_{ijk}^{abc}$ amplitudes are implicit functions of lower-rank 
quantities).

It is important to note that 
various analogues of Eq. (\ref{tuck1}) have also been employed in the CC
calculations for the rank-reduction of the $T_2$ amplitudes tensor (see 
Refs.~\onlinecite{kinoshita03},\onlinecite{benedikt13},\onlinecite{parrish19}
and references therein). In particular, in the recent 
work of Parrish et al.\cite{parrish19} it has been shown that impressive reductions of the computational 
effort can be obtained if the CCSD amplitudes are expanded in a set of highest magnitude 
eigenvectors of the MP2 or MP3 doubles amplitudes. From a mathematical point of view, this idea 
is very similar to the formalism employed in this work for the $T_3$ amplitudes based on Eq. 
(\ref{tuck1}).

Generally speaking, tensor decomposition methods have a long history in quantum chemistry, although 
many of them were introduced with help of a different formalism and employing a different language. 
Most of the effort, until quite recently, has been concentrated on reducing storage requirements 
and computational effort of handling the electron repulsion integrals (ERI), denoted $(pq|rs)$ 
further in the text (Coulomb notation). Probably the most popular method of compressing the 
ERI tensor is the density fitting (DF) 
approximation\cite{whitten73,dunlap79,vahtras93,feyer93,rendell94,kendall97,weigend02} (also known as the 
resolution-of-identity approximation in the present context) where the integrals are expanded into a 
distinct pre-optimized auxiliary basis set, $Q$, as $(pq|rs)\approx B_{pq}^Q\,B_{rs}^Q$. This 
idea was first proposed in the context of self-consistent field calculations, but since then it 
has spread to the more advanced electronic structure methods such as MP2 or CC 
\cite{epifanovsky13,deprince13,deprince14}. Somewhat later the 
Cholesky decomposition (CD) of the ERI tensor has been put forward as an important 
alternative\cite{beebe77,roeggen86,koch03,auilante07,auilante09}. 
While the decomposition format in CD is effectively the same as in DF, the $Q$ basis set does not 
have to be pre-optimized since it is composed of products of the original basis set 
functions. The main advantage of both CD and DF approximations is that the $Q$ basis 
set can be much smaller than the formal rank of the ERI tensor would suggest without an appreciable loss of accuracy. 
More advanced techniques such as the 
pseudospectral\cite{friesner85,friesner86,friesner87,friesner88,
friesner90a,friesner90b,friesner94,friesner99} (PS) and chain-of-spheres 
exchange\cite{neese09,kossmann10,izsak11,taras11,izsak12,izsak13,dutta16} (COSX) methods approximate the ERI tensor in 
the form 
$(pq|rs)\approx X_p^Q\,Y_q^Q\,V_{rs}^Q$, where the $Q$ summation runs over a set of pre-selected 
grid points in three-dimensional space. Even more sophisticated ERI decomposition formats have been 
proposed recently. For example, the tensor hypercontraction (THC) format\cite{hohenstein12,parrish12,parrish13} 
assuming the form $(pq|rs)\approx X_p^P X_q^P Z_{PQ} X_r^Q X_s^Q$ have been introduced, and Benedikt et 
al.\cite{benedikt11,benedikt13} reported representation of the ERI tensor in canonical product (CP) format.

Tensor decompositions can be applied equally well to other quantities appearing in the electronic structure theory 
besides the electron repulsion integrals. One important example is the Laplace transformation of the energy 
denominators that appear in MP and CC theories, introduced first by Alml\"{o}f\cite{almlof91}. It has also been 
realized that tensor decomposition techniques can not only reduce the memory/disk storage requirements and overall 
computational cost of quantum chemistry methods, but in certain cases they can also decrease their formal scaling 
with the system size, see Refs.~\cite{yousung04,aquilante07,schumacher15,song17} as representative examples. 
However, applications of tensor decomposition techniques to the coupled-cluster amplitudes (at CCSD and higher levels 
of theory) is a relatively novel development. Starting with the pioneering papers of Kinoshita et al. 
\cite{kinoshita03} 
where the higher-order singular value decomposition (HOSVD) was adopted for compression of $T_2$ and $T_3$ amplitudes, 
several decomposition formats were proposed. Benedikt et al. \cite{benedikt13} reported application of canonical 
product format (also known as parallel factor decomposition) to the $T_2$ amplitudes at the CCD level of theory, 
resulting in a substantial scaling reduction down to $N^5$. Quite recently the tensor hypercontraction (THC) 
representation has been applied to the CCSD amplitudes\cite{hohenstein12b,schutski17} and it has been demonstrated that 
this reduces the scaling of CCSD iterations to $N^4$. Finally, we point out that many other techniques that have been 
used 
in the literature to reduce the cost of the electronic structure computations can be analyzed and compared more 
systematically if understood as specific tensor decompositions. Examples include methods like optimized virtual orbital 
space\cite{adam87,adam88,neo05,pitoniak06} (OVOS), frozen natural orbitals\cite{sosa89,taube05,taube08,deprince13} 
(FNO), orbital-specific virtuals\cite{yang11,kura12,yang12,schutz13} (OSV), and even some local correlation treatments 
based on, e.g., local pair natural orbitals\cite{neese09b,riplinger13a,riplinger13b,liakos15,schwilk17} (LPNO).

\section{Preliminaries}
\label{sec:prelim}

\subsection{Coupled cluster theory}
\label{sec:cc}

In this work we adopt a particular convention regarding the indices appearing in the expressions. A detailed 
explanation is given in Table \ref{tab:notation}. The canonical Hartree-Fock determinant, denoted $|\phi_0\rangle$, is 
assumed as the reference wavefunction and the spin-orbital energies are given by $\epsilon_p$. The usual partitioning 
of the electronic Hamiltonian, $H=F+W$, into the sum of the total Fock operator ($F$) and the fluctuation potential 
($W$) is used throughout this paper. The singly, doubly, triply, etc., excited state determinants are denoted by 
$|_i^a\rangle$, $|_{ij}^{ab}\rangle$, $|_{ijk}^{abc}\rangle$, and so forth.  For further use we also 
introduce the following conventions: $\langle A\rangle \stackrel{\mbox{\tiny def}}{=} \langle \phi_0 | A \phi_0 
\rangle$, $\langle_i^a|\, A\rangle \stackrel{\mbox{\tiny def}}{=} \langle_i^a | A \phi_0 \rangle$, and 
$\langle A|B\rangle \stackrel{\mbox{\tiny def}}{=} \langle A \phi_0|B \phi_0 \rangle$ for arbitrary operators $A$, $B$. 
All equations reported in this work were derived for closed-shell systems in the spin-restricted formalism.

\begin{table}[t]
  \caption{Details of the notation adopted in the present work; $O$ is the number of active (correlated) orbitals 
  occupied in the reference, $V$ is the number of virtual orbitals.}
  \label{tab:notation}
  \begin{tabular}{llc}
    \hline
    Indices  & Meaning & Range  \\
    \hline
    $i$, $j$, $k$, $l,\ldots$ & active orbitals occupied in the reference      & $O$ \\
    $a$, $b$, $c$, $d,\ldots$ & orbitals unoccupied in the reference (virtual) & $V$ \\
    $p$, $q$, $r$, $s,\ldots$ & general orbitals (occupation not specified)    & $N$  \\
    $\mu$, $\sigma$, $\lambda$, $\nu,\ldots$ & one-particle (atomic) basis set & $N$ \\
    $P$, $Q,\ldots$           & density fitting auxiliary basis set            & $N_{\mathrm{aux}}$ \\
    $X$, $Y$, $Z,\ldots$      & compressed subspace of the triply excited amplitudes & $N_{\mathrm{SVD}}$ \\
    \hline
  \end{tabular}
\end{table}

In the coupled cluster theory the electronic wavefunction is parameterized by an exponential Ansatz
\begin{align}
\label{cc1}
 |\Psi\rangle = e^T\,|\phi_0\rangle,
\end{align}
where $T=\displaystyle\sum_{n=1} T_n$ is the cluster operator that depends on the single ($t_i^a$), double 
($t_{ij}^{ab}$), triple ($t_{ijk}^{abc}$), etc., excitation amplitudes as defined in Ref.~\onlinecite{bartlett07}.
The coupled-cluster equations that are used to 
determine the amplitudes are obtained by inserting 
the Ansatz (\ref{cc1}) into the electronic Schr\"{o}dinger equation, multiplying by $e^{-T}$ from the left, and 
projecting onto a proper subset of excited state determinants. This gives rise to the coupled-cluster 
residual tensors
\begin{align}
 R_i^a = \langle_i^a|\,e^{-T} H e^{T} \rangle,\;\;\; R_{ij}^{ab} = \langle_{ij}^{ab}|\,e^{-T} H e^{T} \rangle,\;\;\;
\mbox{etc}.
\end{align}
They are exactly zero for converged amplitudes, but must be re-calculated a number of times during the 
coupled-cluster iterations. In this work we are concerned with the CCSDT theory where the cluster operator is truncated 
at the level of triples, i.e., $T=T_1+T_2+T_3$. The corresponding residual tensors can be written as
\begin{align}
 &R_i^a = R_i^a(\,\mbox{ccsd}) + \langle_i^a|\left[ W, T_3 \right]\rangle, \\
 &R_{ij}^{ab} = R_{ij}^{ab}(\,\mbox{ccsd}) + \langle_{ij}^{ab}|\left[ \widetilde{F} + \widetilde{W}, T_3 \right]\rangle,
\end{align}
where $R_i^a(\,\mbox{ccsd})$ and $R_{ij}^{ab}(\,\mbox{ccsd})$ originate from the CCSD theory and depend 
only on $T_1$ and $T_2$ (explicit expressions can be found, for example, in Refs.~\onlinecite{koch94,koch96}), and
\begin{align}
\label{r3}
\begin{split}
 R_{ijk}^{abc} &= \langle_{ijk}^{abc}|\left[\widetilde{F},T_3\right]\rangle + 
\langle_{ijk}^{abc}|\left[\widetilde{W},T_2\right]\rangle + 
\half\,\langle_{ijk}^{abc}|\left[\left[\widetilde{F}+\widetilde{W},T_2\right],T_2\right]\rangle \\
 &\;+ \langle_{ijk}^{abc}|\left[\widetilde{W},T_3\right]\rangle +
 \langle_{ijk}^{abc}|\left[\left[W,T_2\right],T_3\right]\rangle.
\end{split}
\end{align}
Throughout the present work we employ the $T_1$-similarity-transformed formalism based on the 
operators $\widetilde{F}=e^{-T_1} F 
e^{T_1}$ and $\widetilde{W}=e^{-T_1} W e^{T_1}$. Correspondingly, the ``dressed'' two-electron integrals, defined 
precisely in Ref.~\onlinecite{koch94}, are denoted by $(pq\widetilde{|}rs)$. The main advantage of this formalism is 
that the 
singly-excited amplitudes can be absorbed into the two-electron integrals, thereby reducing the length of the working 
equations significantly. While this 
comes at a price of performing 
four-index integral transformation during every coupled-cluster iteration, the cost of this procedure is marginal when 
the density fitting approximation is in effect. Explicit expressions for the CCSDT residuals given in terms of basic 
one- and two-electron integrals, and the cluster amplitudes are known in the literature~\cite{noga87}. For closed-shell 
systems one obtains in the present notation
\begin{align}
\label{r1}
 R_i^a = R_i^a(\,\mbox{ccsd}) + \left[ 2(jb|kc) - (jc|kb) \right] \left( t_{ijk}^{abc} - 
 t_{ijk}^{bac} \right),
\end{align}
and
\begin{align}
\label{r2}
\begin{split}
 R_{ij}^{ab} = R_{ij}^{ab}(\,\mbox{ccsd}) &+ P_2\Big[ \widetilde{F}_{kc} \left( t_{ijk}^{abc} - 
 t_{ijk}^{acb} \right) \\ 
 &+ (ac\widetilde{|}kd) \left( 2t_{ijk}^{cbd} - t_{ijk}^{cdb} - t_{ijk}^{dbc}  \right) \\ 
 &- (ki\widetilde{|}lc) \left( 2t_{ljk}^{cba} - t_{ljk}^{bca} - t_{ljk}^{abc} \right) \Big],
\end{split}
\end{align}
where $P_2 = \left(\,_{ij}^{ab}\right) + \left(\,_{ji}^{ba}\right)$. Moving to the triples residual 
let us first define ``long'' and ``short'' permutation operators as
\begin{align}
\label{plong}
 &\mathcal{P}_L = \left(\,_{ijk}^{abc}\right) + \left(\,_{ikj}^{acb}\right) + 
\left(\,_{jik}^{bac}\right) + \left(\,_{kij}^{cab}\right) + 
\left(\,_{jki}^{bca}\right) + 
\left(\,_{kji}^{cba}\right),\\
\label{pshort}
 &\mathcal{P}_S = \left(\,_{ijk}^{abc}\right) + \left(\,_{jik}^{bac}\right) + \left(\,_{kji}^{cba}\right),
\end{align}
respectively, where the notation for the orbital index permutations is the same as in 
Ref.~\onlinecite{koch97}, for example, $\left(\,_{ikj}^{acb}\right)$ permutes the compound indices 
$bj$ and $ck$, while $\left(\,_{jki}^{bca}\right)$ permutes $bj$ and $ck$, and then $ai$ and $bj$.
This allows us to write
\begin{align}
\label{r3b}
\begin{split}
 R_{ijk}^{abc} = \mathcal{P}_L \Big[ t_{il}^{ab}\,\Xi_{ck}^{lj}-t_{ij}^{ad}\,\Xi_{ck}^{bd} \Big] &+ 
 \mathcal{P}_S \Big[ \chi_{li}\,t_{ljk}^{abc} - \chi_{ad}\,t_{ijk}^{dbc} - \chi_{lj}^{mk}\,t_{ilm}^{abc}
 - \chi_{bd}^{ce}\,t_{ijk}^{ade} \\
 &+ \chi_{ad}^{li}\,t_{ljk}^{dbc} + \chi_{bd}^{li}\,t_{ljk}^{adc} + \chi_{cd}^{li}\,t_{ljk}^{abd}
 - \chi_{ai}^{ld} \left( 2t_{ljk}^{dbc} - t_{ljk}^{cbd} - t_{ljk}^{bdc} \right) \Big],
\end{split}
\end{align}
where a handful of intermediates have been introduced
\begin{align}
\label{chi1}
 &\chi_{li} = \widetilde{F}_{li} + (me|ld)\,\bar{t}_{im}^{de},\;\;\;
  &&\chi_{ad} = \widetilde{F}_{ad} - (me|ld)\,\bar{t}_{lm}^{ae},\\
\label{chi2}
 &\chi_{lj}^{mk} = (lj\widetilde{|}mk) + (ld|me)\,t_{jk}^{de},\;\;\;
  &&\chi_{bd}^{ce} = (bd\widetilde{|}ce) + (ld|me)\,t_{lm}^{bc},\\
\label{chi3}
 &\chi_{ad}^{li} = (ad\widetilde{|}li) - (le|md)\,t_{mi}^{ae},\;\;\;
  &&\chi_{ai}^{ld} = (ai\widetilde{|}ld) - (le|md)\,t_{im}^{ae} + (ld|me)\,\bar{t}_{im}^{ae},
\end{align}
where the symbol $\bar{t}_{ij}^{ab}$ is a shorthand notation for $\bar{t}_{ij}^{ab}=2t_{ij}^{ab}-t_{ij}^{ba}$, and
\begin{align}
\label{xioo}
\begin{split}
 \Xi_{ck}^{lj} &= (ck\widetilde{|}lj) + (lj\widetilde{|}md)\,\bar{t}_{mk}^{dc}
 - (ld\widetilde{|}mj)\,t_{mk}^{dc} - (ld\widetilde{|}mk)\,t_{mj}^{cd} \\
 &+ (cd\widetilde{|}le)\,t_{kj}^{de} + (ld|me) \left( 2t_{mkj}^{ecd} - t_{mkj}^{ced} - 
 t_{mkj}^{dce} \right),
\end{split}
\end{align}
\begin{align}
\label{xivv}
\begin{split}
 \Xi_{ck}^{bd} &= (ck\widetilde{|}bd) - \widetilde{F}_{ld}\,t_{lk}^{bc} + 
 (lk\widetilde{|}md)\,t_{lm}^{cb} + (bd\widetilde{|}le)\,\bar{t}_{lk}^{ec}
 - (be\widetilde{|}ld)\,t_{lk}^{ec} \\
 &- (ld\widetilde{|}ce)\,t_{lk}^{be} -(ld|me)\left( 2t_{mkl}^{ecb} - t_{mkl}^{ceb} - 
 t_{mkl}^{bce} \right).
\end{split}
\end{align}
Let us analyse the computational costs of evaluating the CCSDT residual tensor. The $R_i^a$ and 
$R_{ij}^{ab}$ residuals scale as $O^3V^3$ and $O^3V^4$ in the leading-order and thus they do 
not constitute a bottleneck in the ordinary CCSDT calculations. Concerning the triples residual, 
computation of all intermediates denoted by the letter $\chi$ scales as $N^6$ (or less) with the 
size of the system. The most expensive among these intermediates is $\chi_{bd}^{ce}$ and its 
computational costs ($O^2V^4$) are very similar to the so-called ``particle-particle ladder 
diagram'' from the CCSD theory. The $\Xi_{ck}^{lj}$ and $\Xi_{ck}^{bd}$ intermediates are 
more expensive with the leading-order terms scaling as $O^4V^3$ and $O^3V^4$, respectively, due to 
presence of triply excited amplitudes tensor in the last term of 
Eq.~(\ref{xioo})~and~Eq.~(\ref{xivv}). However, the most problematic terms are present in the triples 
residual tensor, Eq.~(\ref{r3b}). The terms in the first square bracket and the first two terms in the second square 
bracket scale as $N^7$ while all of the remaining 
ones as $N^8$. The most expensive amongst the latter is the diagram involving five virtual 
indices, $\chi_{bd}^{ce}\,t_{ijk}^{ade}$, which thus scales as $O^3V^5$ and is typically the 
bottleneck in CCSDT calculations for larger systems. An important conclusion of this analysis is 
that if the cost of the CCSDT method is to be reduced to the level of $N^6$, all $\chi$ 
intermediates, defined by Eqs.~(\ref{chi1})-(\ref{chi3}), may be calculated as they stand but the 
scaling of all remaining terms in Eq.~(\ref{r3b}) must be reduced. Let us also note that in 
the conventional CCSDT calculations there is no way to avoid storing the triples residual tensor 
($R_{ijk}^{abc}$) in core memory which incurs roughly $\frac{1}{6}O^3V^3$ memory cost.

\subsection{Decomposition of the integrals}
\label{sec:df}

To decompose the electron repulsion integrals tensor, $(pq|rs)$, the following symmetric formula is used in this work
\begin{align}
 \label{dfint}
 (pq|rs) = B_{pq}^Q\,B_{rs}^Q,
\end{align}
where $Q$ is some auxiliary basis set (ABS). Analogous notation is employed for the ``dressed'' two-electron integrals, 
i.e., $(pq\widetilde{|}rs)=\widetilde{B}_{pq}^Q\,\widetilde{B}_{rs}^Q$. Note that 
$(ia\widetilde{|}jb)=(ia|jb)$ and thus $\widetilde{B}_{ia}^Q=B_{ia}^Q$.

The generic formula (\ref{dfint}) encompasses two the most popular approximations -- the density 
fitting and the Cholesky decomposition. These two 
techniques differ only in the way of selecting the expansion basis, $Q$. In DF method this basis is carefully 
pre-optimized for each atom and orbital basis set combination. Since ABS functions obtained in this way are not 
orthogonal, the expansion coefficients are calculated as
\begin{align}
 B_{pq}^Q = (pq|P)\,[\mathbf{V}^{-1/2}]_{PQ},
\end{align}
where $(pq|P)$ and $V_{PQ}=(P|Q)$ are the three-centre and two-centre electron repulsion integrals, respectively (see 
Ref.~\onlinecite{katouda09} for more precise definitions). By contrast, in the CD approach the auxiliary basis is 
composed of pairs 
of functions from the original (orbital) basis set. Therefore, this technique requires no additional external input; 
moreover, it is easier to control the accuracy. In our DF-CC implementation the ``dressed'' 
integrals are formed directly from the DF representation of the two-electron integrals in the AO 
basis during every coupled-cluster iteration.

From the point of view of the present work it is critical to note that both in DF and CD approximations the size 
of the ABS ($N_{\mathrm{aux}}$) scales linearly with the size of the 
system. This is obvious in case of the DF technique where the ABS for a molecule is simply a union of 
auxiliary basis sets of the constituting atoms. One can thus write $N_{\mathrm{aux}}=c_{\mathrm{aux}}\,N$ 
and, in practice, we have $c_{\mathrm{aux}}\in2-5$. In the CD approach the asymptotic linear scaling of 
$N_{\mathrm{aux}}$ was demonstrated numerically~\cite{koch03}, but somewhat larger values of 
$c_{\mathrm{aux}}$ may be needed to reach the accuracy levels characteristic for DF~\cite{deprince13,epifanovsky13}.

Let us stress that the application of the decomposition (\ref{dfint}) alone does not improve the overall scaling of the 
CCSDT method. As an example consider the term $\chi_{bd}^{ce}\,t_{ijk}^{ade}$. The intermediate $\chi_{bd}^{ce}$ does 
not factorize naturally to the form analogous to Eq. (\ref{dfint}) because of the second term in Eq. (\ref{chi2}). 
Moreover, even if such factorization was forced, e.g., by performing the Cholesky decomposition 
$\chi_{bd}^{ce}=L_{bd}^Q\,L_{ce}^Q$ during every iteration, the resulting working expression, 
$L_{bd}^Q\left(L_{ce}^Q\,t_{ijk}^{ade}\right)$, would still require $N^8$ computational effort to evaluate.

\subsection{Decomposition of the $T_3$ amplitudes}
\label{sec:dt3}

The most troublesome issue related to the Tucker-3 compression format, Eq.~(\ref{tuck1}), is the 
necessity to 
compute the basic expansion tensors, $U^X_{ai}$. For convenience of the readers we briefly summarize 
the optimal 
strategy to find these quantities. Assume that are given some approximate triples amplitude tensor. 
In the present work we employ the following formula
\begin{align}
\label{t32}
 \,^{(2)}t_{ijk}^{abc} = (\epsilon_{ijk}^{abc})^{-1} \langle \,_{ijk}^{abc} | \big[ \widetilde{W}, 
T_2 
\big]\rangle,
\end{align}
where the $T_1$ and $T_2$ amplitudes are taken from the CCSD theory. The superscript~``$(2)$'' was 
added to distinguish 
Eq.~(\ref{t32}) from the exact CCSDT amplitudes tensor and signifies that this formula is accurate 
only through the 
second order in perturbation theory. Eq.~(\ref{t32}) is obtained by retaining only the first two 
terms of 
Eq.~(\ref{r3}), similarly as in the CC3 model. Compared to our previous work~\cite{lesiuk19} the 
``dressed'' 
fluctuation potential 
$\widetilde{W}$ is used in Eq.~(\ref{t32}) instead of $W$. This is due to a unique role of single 
excitations as 
approximate orbital relaxation parameters which may help to improve the results in quasi-degenerate 
situations. Other 
than that, the results obtained with $W$ and $\widetilde{W}$ should be very close, as are their 
computational costs. 
Let us stress that the choice given by Eq.~(\ref{t32}), although natural and self-contained, is 
arbitrary and 
the method presented in this work is applicable also for other sources of approximate triple 
amplitudes.

Having the approximate tensor $^{(2)}t_{ijk}^{abc}$ we perform its ``flattening'', i.e., rewrite it 
as a rectangular 
matrix with dimensions $O^2V^2\times OV$, giving $^{(2)}t_{aibj,ck}$. Next, the singular-value 
decomposition of this 
matrix is performed. The left singular vectors are discarded while the right singular vectors 
constitute 
the desired basis, $U_{ai}^X$. To obtain the optimal compression of the full tensor $t_{ijk}^{abc}$ 
to a 
desired size $N_{\mathrm{SVD}}$, one retains only those vectors $U_{ai}^X$ that correspond to the 
largest singular values of the ``flattened'' matrix (note that the singular values are non-negative 
real numbers). Unfortunately, if a complete singular value decomposition of the $^{(2)}t_{aibj,ck}$ 
were to be 
performed (and subsequently the insignificant singular values/vectors were simply dropped), the 
computational cost of 
the procedure would scale as $N^8$. To avoid this we have recently introduced a technique based on 
Golub-Kahan 
bidiagonalization~\cite{golub65} which enables to selectively find a predefined number of singular 
vectors of the 
matrix 
$^{(2)}t_{aibj,ck}$ that correspond to the \emph{largest} singular values. Since in practice only a 
small number of 
singular vectors (compared with the dimension of the full $t_{ijk}^{abc}$ tensor) is needed in 
Eq.~(\ref{tuck1}) to 
obtain a decent accuracy, significant time savings are achieved. The method is composed mostly of 
left/right 
multiplications of the matrix $^{(2)}t_{aibj,ck}$ by some trial vectors, and thus its formal 
scaling 
is proportional to $N_{\mathrm{SVD}}\,O^3V^3 \propto N^7$. However, in comparison with the previous 
work we managed 
to exploit the fact that the 
sparsity of the $^{(2)}t_{aibj,ck}$ tensor also increases with the system size. A combination of 
screening techniques 
with sparse matrix-vector multiplication routines have enabled us to reduce the computational effort 
of the procedure 
considerably. In practical applications we observed numerically that the effective scaling of the 
method is 
usually in-between $N^6$ and $N^7$ for larger molecules, as demonstrated further in the 
paper, despite the more pessimistic theoretical estimate. Nevertheless, the formal scaling of this 
step is $N^7$ in the worst-case scenario.

The success of the Tucker-3 compression format, Eq.~(\ref{tuck1}), is based on the fact that the dimension of the 
compressed 
tensor $t_{XYZ}$ can be made significantly smaller than the of the $t_{ijk}^{abc}$ tensor ($OV$) without sacrificing 
much 
of accuracy. To quantify the rate of the compression on a relative basis let us define the \emph{compression factor} as 
$\rho=N_{\mathrm{SVD}}/OV$. Clearly, one has $\rho\leq1$ and the results become exact when $\rho\rightarrow1$. One 
of the most important aspects of Eq.~(\ref{tuck1}) is the scaling of $N_{\mathrm{SVD}}$ with the system size. Let 
us first consider the case when both $O$ and $V$ are simultaneously increased. To maintain a constant relative accuracy 
in the correlation energy the value of $N_{\mathrm{SVD}}$ must increase only linearly with the system 
size, i.e., $N_{\mathrm{SVD}}\propto N$. This has been demonstrated recently at the CC3~\cite{koch97} level of theory 
for chains of 
beryllium atoms with increasing length~\cite{lesiuk19}, and in the present work we provide further numerical evidence 
by considering 
a more chemically appealing example of linear alkanes. It is also worth mentioning that virtually the same conclusion 
regarding the dimension of the compressed tensor has recently been reported by Parrish et al.\cite{parrish19} at the 
CCSD 
level of 
theory. They have considered a compression of $t_{ij}^{ab}$ in the form $t_{ij}^{ab}=t_{XY}\, U_{ai}^X\,U_{bj}^Y$, 
where $U_{ai}^X$ are the eigenvectors of MP2 or MP3 amplitudes corresponding to the eigenvalues of the largest 
magnitude, and proven that to maintain a constant relative 
accuracy in the correlation energy, the dimension of the tensor $t_{XY}$ must scale only 
linearly with the system size. Therefore, this appears to be a more general conclusion that may be equally valid for 
analogues of Eq.~(\ref{tuck1}) in higher dimensions. Additionally, we must stress that while the 
quantity $\rho$ is useful in illustrating the compression rate obtained for a given system (for 
example, with different basis sets) it is not transferable between molecules of different size. In 
fact, because $N_{\mathrm{SVD}}$ scales linearly with the system size, $\rho$ decreases and 
eventually vanishes as the system size grows.

It is also important to discuss the scaling of $N_{\mathrm{SVD}}$ in a different case -- the value of $O$ is fixed and 
only $V$ is increased. This corresponds to a situation where, e.g., one performs calculations for the same system 
increasing only the basis set size. It has been shown~\cite{lesiuk19} that in such case the optimal value of $\rho$ 
that maintains a constant relative accuracy decreases, albeit rather slowly. For example, for a set of a dozen or so 
small molecules 
considered in Ref.~\onlinecite{lesiuk19} the average optimal $\rho$ was found to be 12.5\% for cc-pVDZ basis set and 
decreased to 9.2\% for 
cc-pVQZ. Therefore, $N_{\mathrm{SVD}}$ scales sub-linearly with $V$ but the exact scaling is difficult to quantify and 
may depend on the basis set family, presence of linear dependencies, and numerous other factors.

\section{Theory}
\label{sec:theory}

\subsection{Overview}

In this section we present detailed working equations of an approximate CCSDT method where  
single and double excitations are treated in a conventional way while the compression 
given by Eq.~(\ref{tuck1}) is employed for the triply excited amplitudes. For brevity we 
shall refer to this method as SVD-CCSDT in the remainder of the text.

In methods that employ rank-reduction techniques to reduce its computational burden it is critical that the 
calculations 
are performed without ``unpacking'' the compressed quantities to its original dimension at any stage. 
Therefore, in the SVD-CCSDT method the $t_{ijk}^{abc}$ tensor never 
appears explicitly and is replaced by its compressed counterpart, $t_{XYZ}$. Similarly, instead of 
evaluating the triples residual $R_{ijk}^{abc}$, only the following compressed 
quantity
\begin{align}
\label{rxyzdef}
 r_{XYZ} = U_{ai}^X\,U_{bj}^Y \,U_{ck}^Z \,R_{ijk}^{abc},
\end{align}
is exploited in the calculations. Note that $r_{XYZ}$ is fully symmetric with respect to exchange 
of all its indices. As a byproduct of the SVD procedure the tensors $U_{ai}^X$ form an orthonormal 
basis
\begin{align}
\label{uai1}
 U^X_{ai}\, U^Y_{ai} = \delta_{XY}.
\end{align}
While not strictly necessary it is also beneficial to follow Ref.~\onlinecite{hino04} and enforce the 
relationship
\begin{align}
\label{uai2}
 U^X_{ai}\, U^Y_{ai}\, (\epsilon_i-\epsilon_a) = \epsilon_X\,\delta_{XY},
\end{align}
where $\epsilon_X$ are real-valued constants, which is achieved by a unitary rotation among the 
original $U^X_{ai}$ tensors. This leads to a simple prescription for an update of the compressed 
triple excitation amplitudes 
\begin{align}
 \frac{r_{XYZ}}{\epsilon_X+\epsilon_Y+\epsilon_Z} \rightarrow t_{XYZ}.
\end{align}
In other words, during every coupled cluster iteration the residual tensor $r_{XYZ}$ divided by 
the denominator $\epsilon_X+\epsilon_Y+\epsilon_Z$ is added to the ``old'' compressed triples 
tensor. This obviously leads to convergence when the CC iterations are solved, i.e. when 
$r_{XYZ}=0$.

An extensive justification of an analogous scheme at the CCSD level of theory has been given in the work of 
Parrish et al.~\cite{parrish19}  where a Langrangian form of the coupled-cluster equations is used to define a 
suitable 
stationary condition allowing for optimization of the amplitudes. This formalism is straightforward to generalize to 
the present case if the triply de-excited component of the $\Lambda$-amplitudes is written in a form analogous to 
Eq.~(\ref{tuck1}). Let us also point out that due to relatively small size of the compressed triple amplitudes tensor 
($N^3$ scaling) several instances of it can be stored simultaneously. This allows to exploit techniques such as direct 
inversion of iterative subspace (DIIS)~\cite{pulay80,scuseria86}, or similar 
methods~\cite{purvis81,ziolo08,ettenhuber15} that accelerate the convergence of the 
coupled-cluster equations, with $r_{XYZ}$ being a natural candidate for the error vector.

\subsection{Evaluation of the $R_i^a$ and $R_{ij}^{ab}$ residuals}

In this section we consider triples contribution to the singles and doubles residuals given by Eqs.~(\ref{r1}) 
and~(\ref{r2}). In the evaluation of these contributions it is convenient to exploit the following intermediates
\begin{align}
 &B_{ia}^{QX} = \widetilde{B}_{ji}^Q\,U_{aj}^X, \\
 &B_{ai}^{QX} = \widetilde{B}_{ab}^Q\,U_{bi}^X,\\
 &B_{ij}^{QX} = B_{ia}^Q\,U_{aj}^X,\\
 &A_X^Q = B_{ia}^Q\,U_{ai}^X.
\end{align}
The storage requirements for the first three intermediates scale as $N^4$, but it is not necessary to read them into 
memory in full at any stage of the computations.
These definitions allow us to rewrite the triples contribution to the singles residual as
\begin{align}
\label{r1fac}
\begin{split}
 \langle_i^a|\left[ W, T_3 \right]\rangle &= U_{ai}^X \left[ t_{XYZ} \left( 2A_Y^Q A_Z^Q - B_{jk}^{QZ}\,B_{kj}^{QY} 
 \right) \right] \\
 &- U_{aj}^Y \left[ 2B_{ji}^{QX}\left( t_{XYZ} A_Z^Q \right) - t_{XYZ} \left( B_{jk}^{QZ} B_{ki}^{QX} \right) \right].
\end{split}
\end{align}
The last term in the above expression is the most expensive, scaling as $O^3 N_{\mathrm{aux}} N_{\mathrm{SVD}}^2 
\propto N^6$. Note that the presence of some of the brackets in the above equation are not  
necessary from the mathematical point of view; they are introduced to underline the order of operations 
that leads to the optimal scaling of the computational costs for the respective terms. The same convention is adopted 
in the remainder of the text.

For the doubles residual one obtains an analogous factorization
\begin{align}
\label{r2fac}
\begin{split}
 &\langle_{ij}^{ab}|\left[ \widetilde{F} + \widetilde{W}, T_3 \right]\rangle =
 P_2\bigg[ U_{ai}^X\,U_{bj}^Y \Big( t_{XYZ} \big( \widetilde{F}_{kc}\,U_{ck}^Z \big) \Big) 
 -U_{ai}^X \big( t_{XYZ}\,U_{bk}^Z \big) \big( \widetilde{F}_{kc}\,U_{cj}^Y \big) \\
 &+ 2\,U_{bj}^Y \Big( \big( B_{ia}^{QZ} - B_{ai}^{QZ} \big) \big( A_X^Q\,t_{XYZ} \big) \Big) 
 - \big( B_{ia}^{QZ} - B_{ai}^{QZ} \big) \Big( \underline{B_{kj}^{QY} \big( U_{bk}^X\,t_{XYZ}} \big) \Big) \\
 &- \big(U_{bj}^Z\,t_{XYZ}\big)\,\Big( \underline{B_{ak}^{QX} B_{ki}^{QY}} - U_{ak}^Y\,\big( B_{ic}^{QX} B_{kc}^Q \big) 
\Big).
\end{split}
\end{align}
The computational costs of evaluating all terms in the above expression scale as $N^5$ (or less)
except for the two underlined terms which scale as $N^6$ or, more precisely, as $N_{\mathrm{SVD}}^2 O^2 V 
N_{\mathrm{aux}}$ in the rate-determining step. In order to roughly compare this with the scaling of the uncompressed 
doubles 
residual ($O^3V^4$ in the leading-order term) we set $N_{\mathrm{SVD}}\approx V$. In practical 
applications the ratio $N_{\mathrm{SVD}}/V$ is only somewhat larger than the unity with double-zeta basis sets, 
and somewhat smaller than the unity with triple-zeta (or better) basis sets. The speed-up in the evaluation of the 
doubles residual is thus proportional to $\frac{N_{\mathrm{aux}}}{OV}$. Finally, the size of the 
auxiliary basis set is typically several times larger than $V$ and the ratio of $2-5$ is a reasonable estimate.
Therefore, a considerable speed-up in evaluation of the doubles residual tensor according to Eq. (\ref{r2fac}) is 
expected only when $O$ is large and this has been observed in calculations reported in the 
next section of this work.

\subsection{Evaluation of the compressed triples residual}

While the ability to calculate the doubles residual at a reduced cost for large systems is certainly advantageous, this 
step does not constitute the bottleneck in the full CCSDT calculations. The true proving ground for the decomposition 
strategy adopted in this work is the evaluation of the compressed triples residual, Eq. (\ref{rxyzdef}). In this 
section we present fully factorized equations that prove that computation of $r_{XYZ}$ can be accomplished with $N^6$ 
cost. To this end, we first define permutation operators analogous to Eqs. (\ref{plong}) and (\ref{pshort}) but acting 
on the indices of the SVD basis
\begin{align}
\label{plongprime}
 &\mathcal{P}_L^\prime = \left(XYZ\right) + \left(XZY\right) + 
\left(YXZ\right) + \left(ZXY\right) + 
\left(YZX\right) + 
\left(ZYX\right),\\
\label{pshortprime}
 &\mathcal{P}_S^\prime = \left(XYZ\right) + \left(YXZ\right) + \left(ZYX\right),
\end{align}
and note that for an arbitrary tensor $A_{ijk}^{abc}$ one has
\begin{align}
U_{ai}^X\,U_{bj}^Y \,U_{ck}^Z \,\Big(\mathcal{P}_L\,A_{ijk}^{abc}\Big) = \mathcal{P}_L^\prime \Big( U_{ai}^X\,U_{bj}^Y 
\,U_{ck}^Z\,A_{ijk}^{abc}\Big),
\end{align}
and similarly for the ``short'' permutation. 

Let us define half-transformed doubles amplitudes as
\begin{align}
 T_{ai}^X = U_{bj}^X\,t_{ij}^{ab},\;\;\;
 S_{ai}^X = U_{bj}^X\,t_{ij}^{ba},\;\;\;
 \bar{T}_{ai}^X = 2\,T_{ai}^X - S_{ai}^X
\end{align}
along with a new class of intermediates
\begin{align}
\label{chi1fact}
\begin{split}
 &\chi_{XY}^{\mathrm{occ}} = U_{al}^X\,\big(\chi_{li}\,U_{ai}^Y\big),\;\;\;
  \chi_{XY}^{\mathrm{vir}} = U_{ai}^X\,\big(\chi_{ad}\,U_{di}^Y\big),\\
 &\chi_{XY}^{\mathrm{mix}} = U_{ai}^X\,\big(\chi_{ad}^{li}\,U_{dl}^Y\big),\;\;\;
  \chi_{ld}^X = \chi_{ai}^{ld}\,U_{ai}^X,
\end{split}
 \end{align}
\begin{align}
\label{chi2fact}
 &\chi_{mk}^{XY} = U_{bj}^X\,\big(\chi_{lj}^{mk}\,U_{bl}^Y\big),\;\;\;
  \chi_{ce}^{XY} = U_{bj}^X\,\big(\chi_{bd}^{ce}\,U_{dj}^Y\big),\;\;\;
 \Pi_{li}^{XY} = U_{bj}^X\,\big(\chi_{bd}^{li}\,U_{dj}^Y\big),
\end{align}
and 
\begin{align}
\label{xiz}
 \Xi_{lj}^Z = \Xi_{ck}^{lj}\,U_{ck}^Z,\;\;\;
 \Xi_{bd}^Z = \Xi_{ck}^{bd}\,U_{ck}^Z.
\end{align}
As discussed above, calculation of all $\chi$ intermediates given by Eqs. (\ref{chi1})-(\ref{chi3}) scales as $N^6$ or 
less. Similarly, in Eqs. (\ref{chi1fact})-(\ref{chi2fact}) none of the the step-wise contractions involve more than 
six indices at the same time and thus can be calculated with the computational effort of at most $N^6$. A more 
challenging problem is the evaluation of the last two intermediates, see Eq. (\ref{xiz}), because calculation of 
$\Xi_{ck}^{lj}$ and $\Xi_{ck}^{bd}$ themselves requires an $N^7$ step. Fortunately, by combining
Eqs. (\ref{xioo}) and (\ref{xivv}) with the integral decomposition (\ref{dfint}), and by manipulating the order of 
multiplications one can show that the intermediate $\Xi_{bd}^Z$ can equivalently be rewritten as
\begin{align}
\label{xivvsvd}
\begin{split}
 \Xi_{bd}^Z &= \widetilde{B}_{bd}^Q\,\big(\widetilde{B}_{ck}^Q\,U_{ck}^Z\big) - \widetilde{F}_{ld}\,T_{bl}^Z
 + \underline{U_{ck}^Z\,\Big( t_{lm}^{cb} \big( \widetilde{B}_{lk}^Q\,B_{md}^Q \big) \Big)}+\widetilde{B}_{bd}^Q\,\big( 
B_{le}^Q\,\bar{T}_{el}^Z \big) \\
 &\,- B_{ld}^Q\,\big( \widetilde{B}_{be}^Q\,T_{el}^Z \big) - B_{ld}^Q\,\Big( t_{lk}^{be}\big( 
U_{ck}^Z\,\widetilde{B}_{ce}^Q\big)\Big)
 -2 \underline{B_{ld}^Q\,\big( U_{bl}^{Z'}\,\big( t_{X'ZZ'}\,A_{X'}^Q\big)\Big)} \\
 &\,+ \underline{B_{ld}^Q\,\Big( U_{bl}^{Z'}\,\Big( B_{me}^Q\,\big( U_{ek}^{Y'}\,\big( t_{X'Y'Z'} \big( U_{cm}^{X'} \,
U_{ck}^Z \big)  \big) \Big) \Big) \Big)} + t_{X'Y'Z}\,\Big( U_{bm}^{Y'} \big( B_{ld}^Q\,\widetilde{B}_{ml}^{QX'} \big) 
\Big).
\end{split}
\end{align}
Calculation of all terms in the above expression scales as $N^5$ or less except for the three 
underlined terms that
contain $N^6$ (outer) leading-order steps scaling as $O^3V^3$, 
$O^2V^2N_{\mathrm{SVD}}N_{\mathrm{aux}}$, and $O^2N_{\mathrm{SVD}}^4$, in the order of 
appearance. There are no terms scaling as $N^7$ or higher. Similarly for the $\Xi_{lj}^Z$ intermediate one obtains
\begin{align}
\label{xioosvd}
\begin{split}
 \Xi_{lj}^Z &= \widetilde{B}_{lj}^Q\,\big(\widetilde{B}_{ck}^Q\,U_{ck}^Z\big) + \widetilde{B}_{lj}^Q\,\big( 
B_{md}^Q\,\bar{T}_{dm}^Z \big) - \widetilde{B}_{mj}^Q\,\big( B_{ld}^Q\,T_{dm}^Z \big) - 
U_{ck}^Z\,\Big( t_{mj}^{cd} \big( B_{ld}^Q\,\widetilde{B}_{mk}^Q \big) \Big) \\
& + U_{ck}^Z\,\Big( t_{kj}^{de} \big( B_{le}^Q\,\widetilde{B}_{cd}^Q \big) \Big)
+ 2 B_{ld}^Q\,\Big( U_{dj}^{Z'}\,\big( t_{X'ZZ'} A_{X'}^Q \big) \Big) \\
&-B_{ld}^Q\,\Big( U_{dj}^{Z'}\,\Big( B_{me}^Q\,\big( U_{ek}^{Y'}\,\big( t_{X'Y'Z'} \big( U_{cm}^{X'} \,
U_{ck}^Z \big)  \big) \Big) \Big) \Big)
+ t_{X'ZZ'} \big( B_{lm}^{QX'}\,B_{mj}^{QZ'} \big).
\end{split}
\end{align}
One may notice that many terms present in Eq. (\ref{xioosvd}) bare close resemblance to analogous terms in Eq. 
(\ref{xivvsvd}). Indeed, in a careful implementation many intermediate quantities necessary for evaluation of 
$\Xi_{bd}^Z$ can be reused when $\Xi_{lj}^Z$ is constructed simultaneously. This allows to compute $\Xi_{lj}^Z$ 
essentially as a byproduct with only a handful of additional terms that need to be evaluated separately. The 
latter terms are relatively inexpensive since they scale as $O^3 N_{\mathrm{SVD}}^2N_{\mathrm{aux}}$ or similarly.

Finally, we pass to the calculation of the compressed triples residual tensor, Eq. (\ref{rxyzdef}). 
With help of the intermediates defined above it can be rewritten as
\begin{align}
\label{rxyzfact}
\begin{split}
 r_{XYZ} &= \mathcal{P}_L^\prime\bigg[ \Big( T_{bl}^X\,\Xi_{lj}^Z - \Xi_{bd}^Z\,T_{dj}^X \Big)U_{bj}^Y\bigg]
 + \mathcal{P}_S^\prime\bigg[ \,\chi_{X'X}^{\mathrm{occ}}\,t_{X'YZ} - \chi_{XX'}^{\mathrm{vir}}\,t_{X'YZ} \\
 &+\chi_{XX'}^{\mathrm{mix}} \Big( t_{X'Y'Z}\,\big(U_{bj}^Y\,U_{bj}^{Y'}\big)\Big) 
  -2\,t_{X'YZ}\,\big(\chi_{ld}^X\,U_{dl}^{X'}\big) \\
 &-U_{ck}^Z\,\Big( \chi_{mk}^{YY'} \big( t_{XY'Z'}\,U_{cm}^{Z'}\big) \Big)
  -\underline{U_{ck}^Z\,\Big( \chi_{ce}^{YY'} \big( t_{XY'Z'}\,U_{ek}^{Z'}\big)} \Big)
   \\
  &+U_{ai}^X\,\Big( U_{al}^{X'}\,\big( t_{X'Y'Z}\,\Pi_{li}^{YY'}+t_{X'YZ'}\,\Pi_{li}^{ZZ'}\big)\Big)
  \\
  &+ U_{ck}^Z\,\Big( \big( t_{X'Y'Y}\,U_{cl}^{X'}\big)\big( \chi_{ld}^X\,U_{dk}^{Y'}\big)\Big)
   + U_{bj}^Y\,\Big( \big( t_{X'Y'Z}\,U_{bl}^{X'}\big)\big( \chi_{ld}^X\,U_{dj}^{Y'}\big)\Big) \bigg].
\end{split}
\end{align}
Computation of the terms in the first square brackets scales as $N^5$ -- less expensive than 
of the $\Xi_{lj}^Z$ and $\Xi_{bd}^Z$ intermediates themselves. The first four terms in the second square brackets scale 
as $N^4$, and thus are not a cause for a major concern, while the remaining terms in Eq. (\ref{rxyzfact}) scale as 
$N^6$. The most 
expensive part of Eq. (\ref{rxyzfact}) is the sixth term in the second square brackets (underlined, scaling as
$OV^2N_{\mathrm{SVD}}^3$) resulting from factorization of the $\chi_{bd}^{ce}\,t_{ijk}^{ade}$ term of the 
uncompressed triples residual, cf. Eq. (\ref{r3b}). This is the only term in the conventional CCSDT method that 
involves five virtual indices simultaneously, leading to a rate-determining computational step that scales as 
$O^3V^5$. Under the assumption that $V\approx N_{\mathrm{SVD}}$ we can therefore conclude that the compression method 
employed in this work reduces the cost of evaluating the CCSDT triples residual by a factor proportional to $O^2$.

In the above discussion we have neglected an important issue. The major difference between the conventional CC 
and SVD-based approaches is the number of consecutive tensor contractions that appear in the working expressions. In 
the 
conventional CCSDT residual, Eq. (\ref{r3b}), no products containing more than three different tensors are present. 
Therefore, there are only three possible ways to perform contractions of the constituting tensors and all necessary 
manipulations can be performed rather easily. In the SVD-based formalism, on the other hand, some quantities contain 
products of six or seven tensors -- see, for example, Eq. (\ref{xivvsvd}). The number of ways the tensors can be 
arranged grows exponentially with the length of the tensor string, and in most cases the order of tensor 
multiplications effects the final scaling of a given term. Since typically it is not possible to 
check all reasonable arrangements by hand, most of the 
derivations presented in this work were accomplished with help of computer algebra\cite{wick} which allowed to 
determine the optimal multiplication order by defining a set of rules. 

An additional problem that appears in this context is related to the fact that while the scaling of the computational 
cost for a given expression is always defined uniquely, 
the prefactors of two terms with the same scaling are, in general, difficult to compare. For 
example, the relative cost of two terms with the same scaling may depend on the difference between the ratios 
$N_{\mathrm{SVD}}/V$ and $V/O$. This difference may change significantly depending on the number of electrons in the 
system, cardinality of the basis set, desired accuracy threshold, etc. In such problematic cases several variants of 
a routine handling the same tensor string should probably be incorporated into the code and the decision which one is 
to be used should be made on the fly. However, in our pilot implementation reported in this work, we simply selected 
those factorizations that delivered a reasonable efficiency in a wide range of situations assuming $O<V\approx 
N\approx N_{\mathrm{SVD}}<N_{\mathrm{aux}}$ as a rule of thumb.

Finally, let us discuss memory requirements of the SVD-CCSDT method in the present implementation and some technical 
aspects of 
handling numerous intermediate quantities that appear in the working equations. The only objects of size $N^4$ 
that must be held in core memory are the double excitation amplitudes and the doubles residual tensor 
(both~$\frac{1}{2}O^2V^2$). 
Other ``large'' intermediates are either stored on the disk and read in smaller chunks whenever necessary 
($B_{ia}^{QX}$, $B_{ai}^{QX}$, and $B_{ij}^{QX}$) or are built on-the-fly in a batched loop over one occupied/virtual 
index ($\chi_{mk}^{XY}$, $\chi_{ce}^{XY}$, $\Pi_{li}^{XY}$) and contracted immidiately. Therefore, there is no need to 
keep any of them in full in core memory. The remaining quantities require only $N^3$ memory to store -- the largest 
being either $\Xi_{bd}^Z$ or $t_{XYZ}$ depending on the circumstances. To sum up, while the overall memory requirements 
of the SVD-CCSDT method are significantly larger than of CCSD, the SVD-CCSDT algorithm introduces no large intermediate 
quantities 
that would create a serious memory bottleneck. In particular, the uncompressed $t_{ijk}^{abc}$ and $R_{ijk}^{abc}$ 
tensors do not appear at any stage of the calculations which removes the $N^6$ memory requirement of the conventional 
CCSDT theory.

\section{Numerical results and discussion}
\label{sec:numer}

\subsection{Computational details}

All calculations reported in this work employ Dunning-type cc-pVDZ and cc-pVTZ basis sets~\cite{dunning89}. The 
corresponding auxiliary basis sets (MP2FIT) for the density fitting approximation 
were taken from the work of Weigend et al.~\cite{weigend98,weigend02} Pure spherical representation of both basis sets 
was used throughout.

The theoretical methods described in the previous sections were implemented in a locally modified version of the 
\textsc{Gamess} program package~\cite{gamess1}. The code for the DF decomposition is based on the of the 
resolution-of-identity MP2 (RI-MP2) implementation by Katouda and Nagase~\cite{katouda09} that is available in the 
official release of the \textsc{Gamess} program. DF was used by default at every stage 
of correlated calculations, i.e., in the remainder of the text the acronyms CCSD, CC3, etc., should be interpreted as 
DF-CCSD, DF-CC3, and so forth, unless explicitly stated otherwise. The only method where DF is never used is the 
uncompressed (exact) CCSDT since, to the best of our knowledge, no such implementation is publicly available. 
Hartree-Fock equations were always solved 
utilising the exact two-electron integrals. Frozen-core approximation was invoked throughout: $1s$ orbitals of 
all first-row atoms were left uncorrelated (inactive) unless explicitly stated otherwise. Geometries of the molecules 
were optimized at MP2/cc-pVTZ level of theory. The only exceptions are linear alkanes, C$_n$H$_{2n+2}$ 
with $n=1,8$, which were optimized by using B3LYP/cc-pVTZ method~\cite{becke92,stephens94,hertwig97}. Geometries of all 
molecules considered 
in this work can be found in Supporting Information. Other parameters controlling the coupled-cluster calculations and 
decomposition steps were the same as in Ref.~\onlinecite{lesiuk19}.

\subsection{Scaling with the system size}

\begin{figure}[t]
 \includegraphics[scale=1.00]{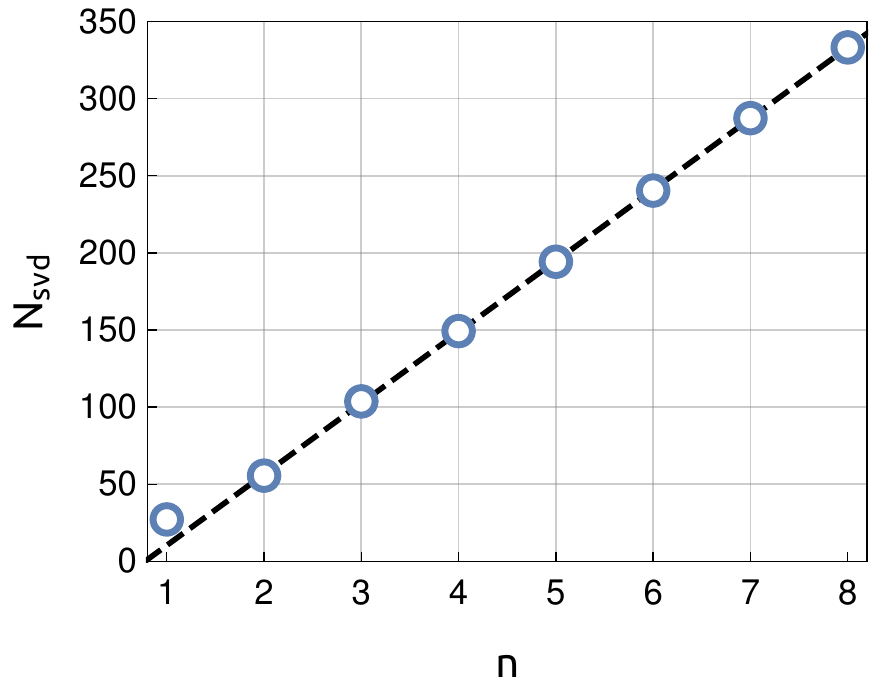}
 \caption{\label{cncc3} Optimal size of the SVD subspace ($N_{\mathrm{SVD}}$) sufficient to recover 99.9\% of the 
 CC3 correlation energy for linear alkanes, C$_n$H$_{2n+2}$, as a function of the chain length  ($n$). }
\end{figure}

Before discussing the accuracy of the SVD-CCSDT method we would like to demonstrate that, in calculations 
for realistic systems, its scaling is consistent with the theoretical findings from the previous section. To this end 
we performed CC3 and SVD-CC3 calculations for linear alkanes, C$_n$H$_{2n+2}$, with the chain length $n=1,\ldots,8$. 
For 
each~$n$ we recorded the optimal size of the SVD subspace, $N_{\mathrm{SVD}}$, sufficient to recover 99.9\% of the 
CC3 correlation energy (uncompressed CC3 was used as a benchmark). The results presented in Fig. \ref{cncc3} 
reveal almost perfect linear relationship between $N_{\mathrm{SVD}}$ and $n$ for $n\geq2$. The coefficient of 
determination for the linear fit to the data with $n\geq2$ is higher than 0.999. Next, we performed SVD-CCSDT 
calculations for the same set of molecules with $N_{\mathrm{SVD}}$ found at the CC3 level of theory for each~$n$. 
Additionally, for $n=2-6$ we performed the exact (uncompressed) CCSDT calculations with the \textsc{AcesII} program 
package. Unfortunately, for $n>6$ a single CCSDT iteration already took more than several days; such calculations would 
not be feasible in practice and were thus abandoned.

\begin{figure}[t]
 \includegraphics[scale=1.00]{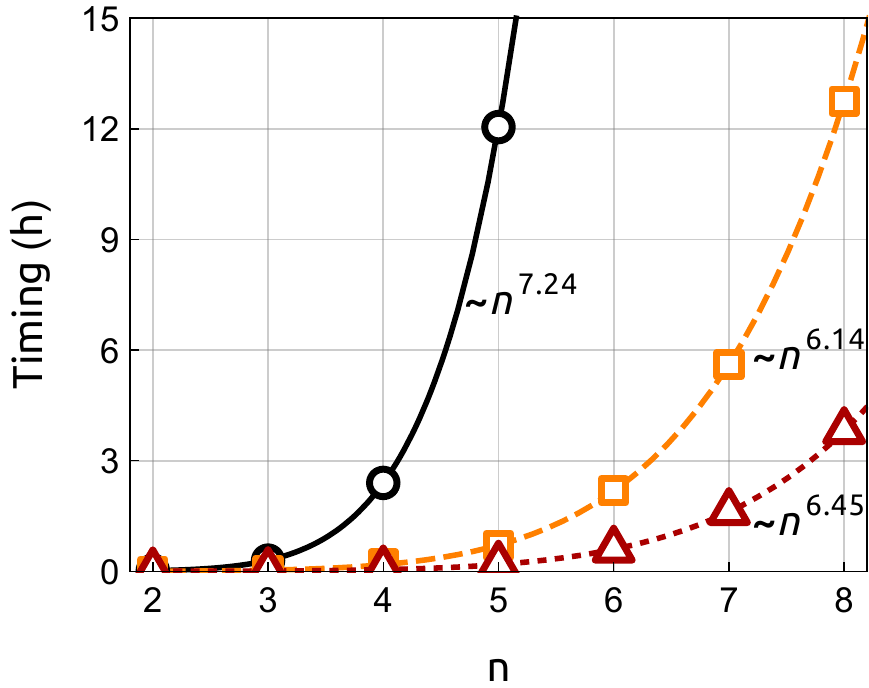}
 \caption{\label{cnscal} Timings for a single iteration of the CCSDT method (solid black line, 
$1.04\cdot10^{-4}\,n^{7.24}$), of the SVD-CCSDT method 
 (dashed orange line, $3.60\cdot10^{-5}\,n^{6.14}$) and of the determination of the SVD subspace 
(dotted red line, $5.75\cdot10^{-6}\,n^{6.45}$) for linear alkanes, 
C$_n$H$_{2n+2}$, as a 
 function of the chain length  ($n$). Power functions obtained by fitting  the correponding data points are given near 
 each graph.}
\end{figure}

In Fig. \ref{cnscal} we compare timings and scaling of the computational costs separately for
\begin{itemize}
 \item the exact CCSDT calculations (solid black line);
 \item SVD-CCSDT calculations with $N_{\mathrm{SVD}}$ determined at the CC3 level of theory (dashed orange line);
 \item determination of the SVD expansion tensors, $U_{ai}^X$ in Eq. (\ref{tuck1}) (dotted red line),
\end{itemize}
taking linear alkanes as a benchmark.
In each case the time necessary for a single iteration is presented (determination of the SVD subspace is also 
iterative in nature, see Ref.~\onlinecite{lesiuk19}). However, it must be noted that the number of iterations necessary 
to converge the 
SVD vectors is usually less than ten, and even five iterations are sufficient for small SVD subspaces. To converge the 
CCSDT and SVD-CCSDT calculations about 20-30 iterations are usually required (albeit we observed that SVD-CCSDT has a 
better convergence characteristics than the exact CCSDT in more demanding cases). Therefore, one has to keep in mind 
that the actual total workload required for the determination of the SVD subspace is, under typical circumstances, 
several times smaller in relation to SVD-CCSDT than the timings for a single iteration would suggest.

The computational timings discussed above were fitted with a power function $a\cdot n^b$ for $n\geq2$, 
where $a$ and $b$ are adjustable real parameters. Since $n$ is roughly proportional to number of orbital and auxiliary
basis set functions, this allows to ``empirically'' determine the scaling of the computational costs with the 
system size. Starting with the CCSDT method, we obtained~$n^{7.24}$ from the fit which is somewhat smaller than the 
expected~$n^8$. This can be explained by the fact that the CCSDT calculations were feasible only up to $n=6$ which 
may not be sufficient to reach the asymptotic regime and lower-order terms may still contribute significantly to the 
total computational time. In the case of the SVD-CCSDT method, where computational timings up to $n=8$ were available, 
the power law fit ($n^{6.14}$) agrees well with the expected scaling of the method. For the determination of 
the SVD expansion tensors the obtained scaling is $n^{6.45}$. This scaling reduction is accompanied by a considerable 
decrease of the overall computational costs. For example, a single iteration of the exact CCSDT method for $n=5$ takes 
a comparable amount of time to a single SVD-CCSDT iteration for $n=8$. By extending the trends shown in Fig. 
\ref{cnscal} one can also estimate that for $n=11-12$ the SVD-CCSDT would reach a cost comparable to CCSDT for 
$n=6$.

\begin{figure}[t]
 \includegraphics[scale=1.00]{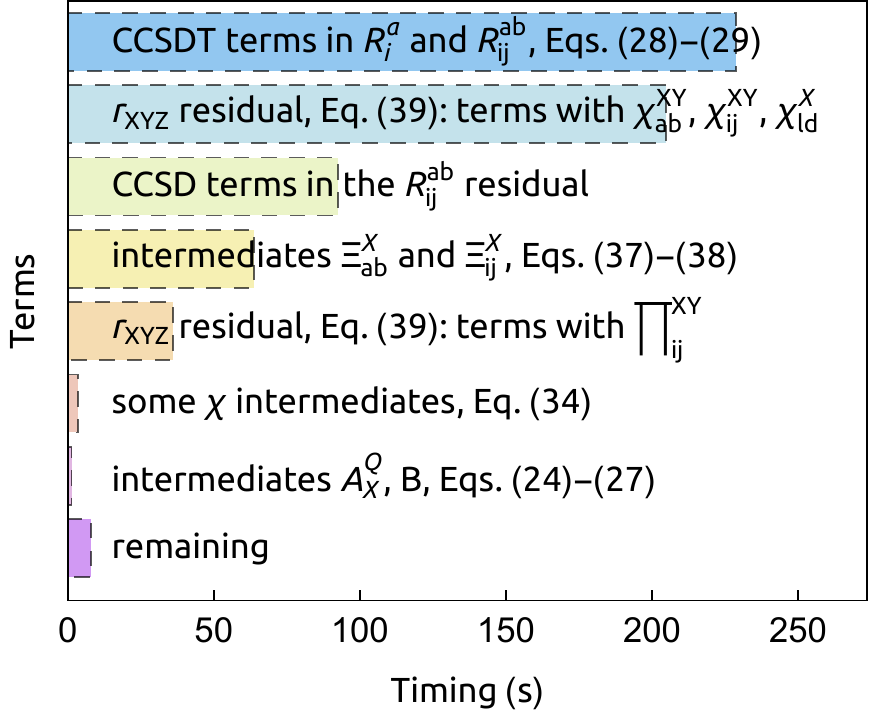}
 \caption{\label{svdterms} Breakdown of the SVD-CCSDT wall clock timing for the HNO$_3$ molecule in 
 the cc-pVTZ basis set ($O=12$, $N=134$, $N_{\mathrm{aux}}=354$, $N_{\mathrm{SVD}}=120$). All steps that required less 
 than 1 second to complete are accumulated in the ``remaining'' category. This category also includes the cost of 
 updating the coupled-cluster amplitudes, convergence checks, evaluating the energy, and similar minor tasks performed 
 during every iteration.}
\end{figure}

It is also interesting to break the computational time spent in a single SVD-CCSDT iteration down into components 
corresponding to various terms in Eq. (\ref{rxyzfact}) and other steps discussed in Sec. \ref{sec:theory}. For this 
purpose we selected HNO$_3$ molecule in the cc-pVTZ basis set ($O=12$, $N=134$, 
$N_{\mathrm{aux}}=354$). As shown in the next section, the size of the SVD subspace considered here 
($N_{\mathrm{SVD}}=120$) is sufficient to reach the accuracy of a fraction of kJ/mol in the total correlation energy 
with respect to the uncompressed result. We thus have $N_{\mathrm{SVD}}\approx V$ which is a typical phenomenon
for the basis sets of this quality. The breakdown of the computational timings is shown in Fig. \ref{svdterms}. 
Rather 
surprisingly, the most computationally demanding step is evaluation of the triples contribution to the singles and 
doubles residuals. The next in the order of expense is one of the terms from Eq. (\ref{rxyzfact}) that involves the 
$\chi_{ab}^{XY}$ intermediate. For comparison, in Fig. \ref{svdterms} we also include the CCSD contribution to the 
$R_{ij}^{ab}$ residual which typically consumes more than 90\% of 
the computational time necessary for the conventional CCSD iterations. This allows to compare the cost of various 
terms present in the SVD-CCSDT theory in relation to the standard CCSD method, revealing the that SVD-CCSDT calculations 
are only $5-6$ times more expensive than CCSD. Since both methods scale as $N^6$ with the size of the 
system, this ratio is likely to remain approximately constant for larger molecules, albeit not necessarily in basis 
sets that are much larger. Let us point out that all calculations reported here were accomplished by using a single CPU 
core. However, it has recently been shown that impressive reductions of the CCSDT computational cost are achievable 
with parallel execution~\cite{prochnow10}. Since our present SVD-CCSDT implementation consist mostly of lengthy loops 
over fixed-size batches of occupied, virtual, etc., indices, we believe that a similar efficiency gain is possible, and 
this option should be considered in future implementations.

\begin{figure}[ht!]
\begin{tabular}{cc}
 \includegraphics[scale=1.00]{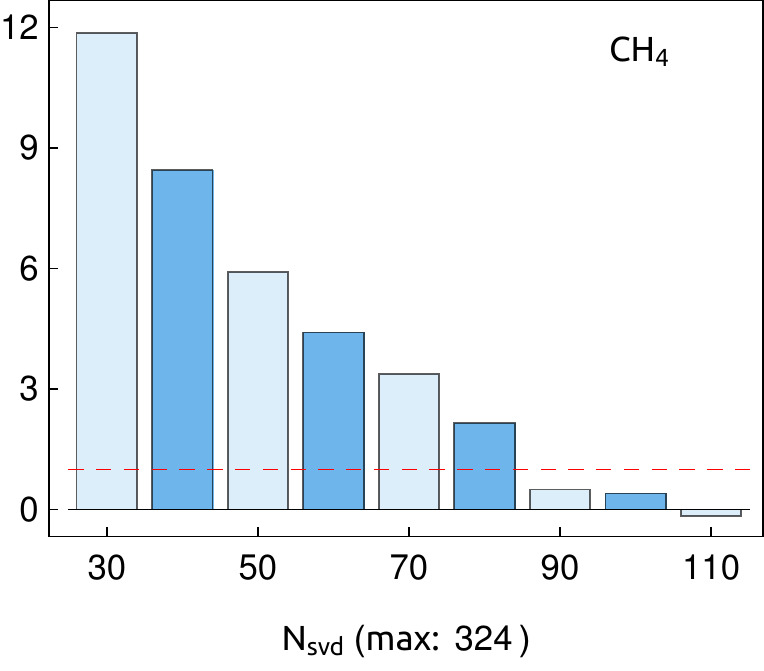} & \includegraphics[scale=1.00]{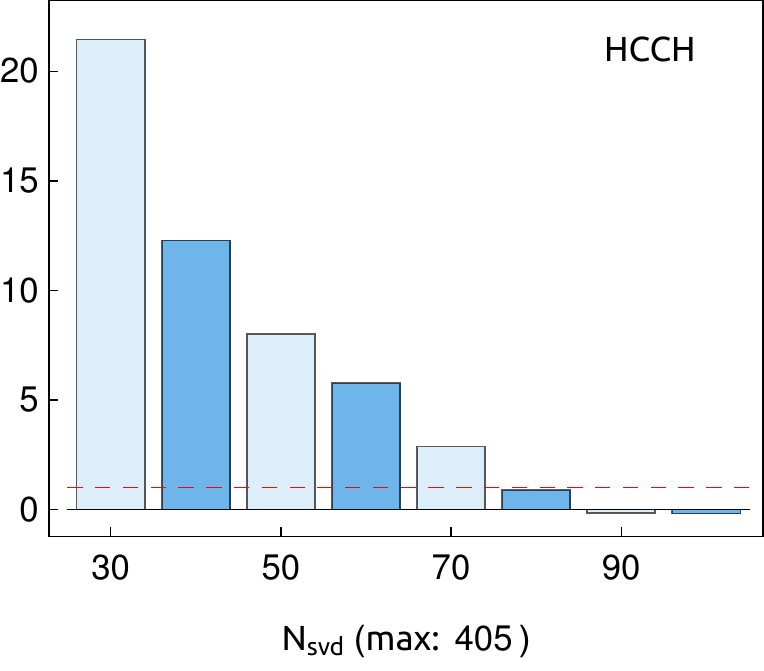} \\
 \includegraphics[scale=1.00]{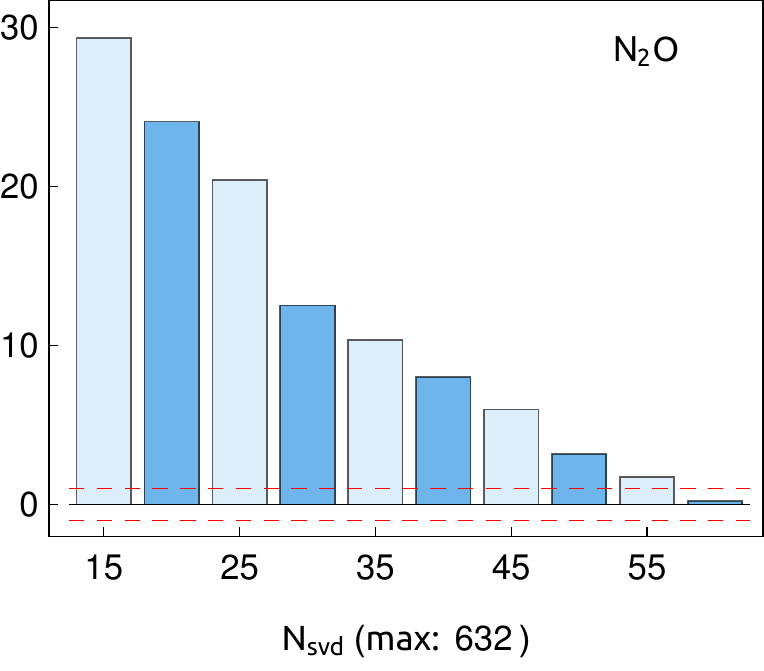} & \includegraphics[scale=1.00]{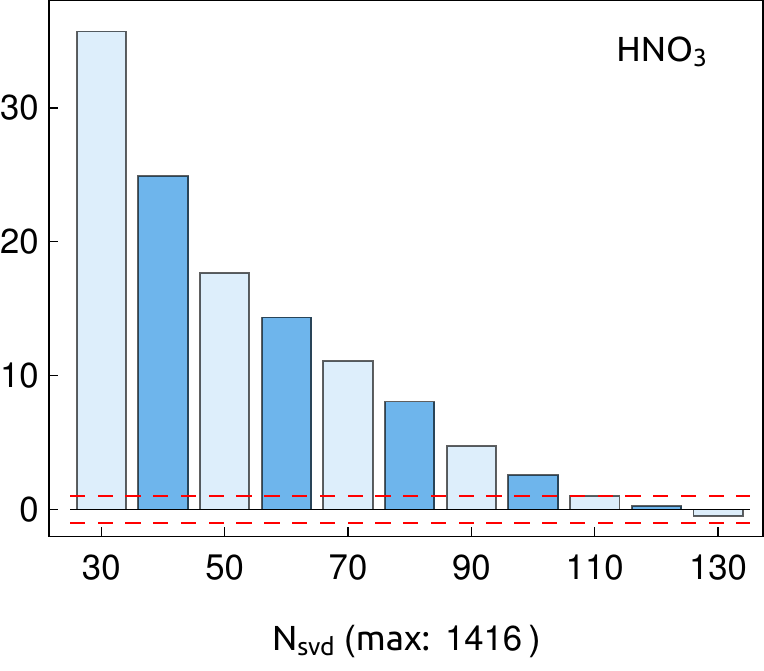} \\
\end{tabular}
 \caption{\label{abserr} Errors in the SVD-CCSDT correlation energies (in kJ/mol) as a function of the SVD subspace 
size ($N_{\mathrm{SVD}}$) with respect to the exact (uncompressed) CCSDT method (cc-pVTZ basis set). The results are 
given for the molecules: methane (upper left panel), ethyne (upper right panel), nitrous oxide (lower left panel), 
and nitric acid (lower right panel). The horizontal red dashed line marks the $1$ kJ/mol accuracy threshold (the 
chemical accuracy). The maximum possible size of the SVD space is given below each graph.}
\end{figure}

\subsection{Accuracy of the method}

To investigate the accuracy of the SVD-CCSDT method in the reproduction of the exact CCSDT energetics we selected four 
small benchmark molecular systems (methane, ethyne, nitrous oxide, and nitric acid) for which the conventional CCSDT 
calculations can be performed in a reasonable wall time. We evaluated SVD-CCSDT energies with increasing SVD 
subspace size (in steps of five at a time) and recorded errors with respect to the exact CCSDT method. Note that 
in the SVD-CCSDT results there is an additional source of error due to the density fitting approximation. To eliminate 
this problem from our benchmark calculations we tested two distinct approaches. The first is to use very large 
auxiliary basis set such as cc-pV6Z-RI (available in the Basis Set Exchange repository~\cite{emsl07}). The second idea 
is to assume that the density-fitting error is the same at the CC3 and CCSDT levels of theory. One can then evaluate 
the 
DF-CC3 and conventional CC3 energies separately and correct the SVD-CCSDT results to account for the difference. In 
all calculations reported here both methods agreed to 0.1 kJ/mol and thus the results reported in this section 
can be viewed as essentially free from the density-fitting error. Of course, this problem occurs only if the total 
correlation energies are compared; in evaluation of relative energies the density-fitting error is known to 
systematically cancel out leaving only a very small residual error.

The results of benchmark calculations are represented graphically in Fig. \ref{abserr} where SVD-CCSDT errors (with 
respect to the conventional CCSDT) are plotted against the SVD subspace size. One can see that the errors vanish rather 
quickly and the assumed 1 kJ/mol accuracy goal (the chemical accuracy) is reached with a small fraction of the total 
number of SVD vectors. The overall picture is very similar to the results reported recently at the CC3 level of theory, 
but we found that there appears to be no systematic relationship connecting the convergence rates of SVD-CC3 and 
SVD-CCSDT methods that would hold for a broader range of systems. At present, the ultimate limit of accuracy of the 
SVD-CCSDT method appears to be the level of $\pm 0.1-0.2$ kJ/mol where oscillations start to appear. This is a 
consequence of increasing numerical instabilities in the procedure of obtaining consecutive SVD vectors and we hope to 
improve the procedures introduced in Ref.~\onlinecite{lesiuk19} to eliminate this problem in future works. Other 
possible strategies to increase the accuracy of the present approach are discussed in the next section. All 
in all, the results presented in Fig. \ref{abserr} indicate that the chemical accuracy of the SVD-CCSDT energies can be 
reached without significant difficulties and even more stringent accuracy levels are obtainable with acceptable sizes of 
the SVD subspace.

\begin{figure}[t]
 \includegraphics[scale=1.00]{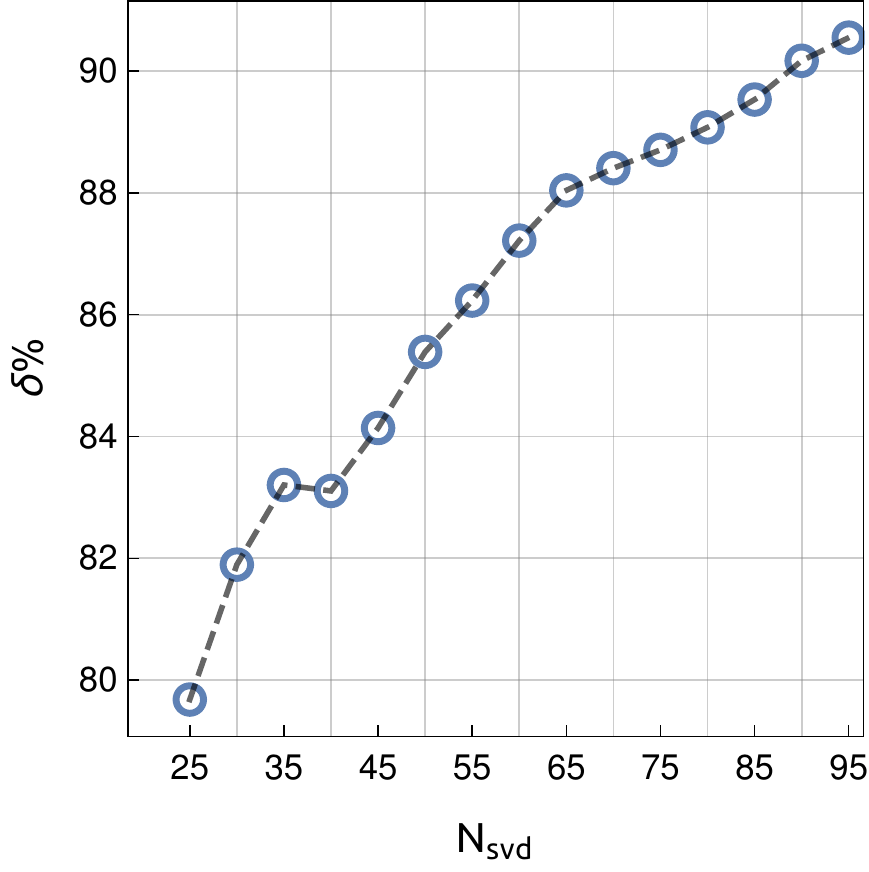}
 \caption{\label{postcc3} Accuracy of the post-CC3 energy contribution, see Eq. (\ref{deltacc3}) for a precise 
 definition, as a function of the SVD subspace size for the nitrous oxide molecule (cc-pVTZ basis set). The black 
 dashed lines are linear functions connecting two neighbouring data points.}
\end{figure}

From the point of view of some applications it would be beneficial to use SVD-CCSDT method to calculate only the 
post-CC3 [or post-CCSD(T)] energy correction, rather than the total SVD-CCSDT result, e.g., when the uncompressed CC3 
results are available (possibly calculated without the density fitting approximation). In this case the post-CC3 
effects can be evaluated as a difference between SVD-CCSDT and SVD-CC3 results obtained with \emph{the same} SVD 
subspace (note that the density-fitting error is cancelled out in the process). To verify whether such approach is 
reasonable let us define the following quantity that measures the accuracy of the calculated post-CC3 effects (on 
percentage basis)
\begin{align}
\label{deltacc3}
 \delta \% = 100\left|\frac{E_{\rm SVD-CCSDT}-E_{\rm SVD-CC3}}{E_{\rm CCSDT}-E_{\rm CC3}}\right|
\end{align}
where $E_{\rm SVD-CCSDT}$, $E_{\rm SVD-CC3}$ are SVD-CCSDT and SVD-CC3 correlation energies, respectively, obtained 
with the same SVD subspace, and $E_{\rm CCSDT}$ and $E_{\rm CC3}$ are the conventional CCSDT and CC3 results. In Fig. 
\ref{postcc3} we illustrate the dependence of $\delta \%$ on the size of the SVD subspace for the the nitrous oxide 
molecule in the cc-pVTZ basis set. Even a very small number of SVD vectors allows to recover about 80\% of the total 
post-CC3 effects. To reproduce about 90\% of the exact value a somewhat larger SVD subspace is required, corresponding 
to $\rho\approx15\%$. A systematic convergence pattern towards the exact value is also notable, but it is not 
necessarily the same as for the raw energies since SVD-CC3 and SVD-CCSDT components may converge at a somewhat 
different rate. Taking into consideration that the calculation of the post-CC3 [or post-CCSD(T)] effects is notoriously 
difficult, as well documented in the literature~\cite{smith14}, the method presented here becomes an interesting 
alternative, especially in larger basis sets and for systems where these effects are essential for achieving the 
chemical accuracy.

\begin{table}[t]
  \caption{Mean absolute deviation and maximum deviation (both in kJ/mol) of the SVD-CCSDT results from the 
corresponding uncompressed CCSDT values for torsional energy in butadiene molecule (cc-pVDZ basis set). The data set 
consists of eighteen CCCC dihedral angles, $\theta=0,10,20,\ldots,170$.}
  \label{tab:buta1}
  \begin{tabular}{cccc}
    \hline
    $N_{\mathrm{SVD}}$  & $\rho$ & mean abs. deviation & maximum deviation  \\
    \hline
    20  & 2.6  & 0.92 & 2.10 \\
    40  & 5.1  & 0.77 & 1.61 \\
    60  & 7.7  & 0.42 & 1.02 \\
    80  & 10.2 & 0.27 & 0.73 \\
    100 & 12.8 & 0.19 & 0.50 \\
    120 & 15.4 & 0.14 & 0.30 \\
    \hline
  \end{tabular}
\end{table}

Finally let us consider calculation of relative energies with the help of SVD-CCSDT method. For this purpose we 
consider the 1,3-butadiene molecule which can assume several interesting geometric structures~\cite{feller09b} that are 
distinguished by the value of the CCCC dihedral angle, $\theta$. The planar \emph{trans} geometry ($\theta=180^\circ$) 
is the global energy minimum while the analogous \emph{cis} structure ($\theta=0^\circ$) is a saddle point on the 
potential energy surface. Interestingly, there is another stable conformer -- the so-called \emph{gauche} structure that 
appears for the dihedral angle $\theta\approx 35^\circ$ and represents a local minimum. The \emph{gauche} and 
\emph{trans} structures are separated by a large energy barrier with a maximum for $\theta\approx 100^\circ$.

\begin{figure}[t!h!]
\begin{tabular}{c}
 \includegraphics[scale=1.00]{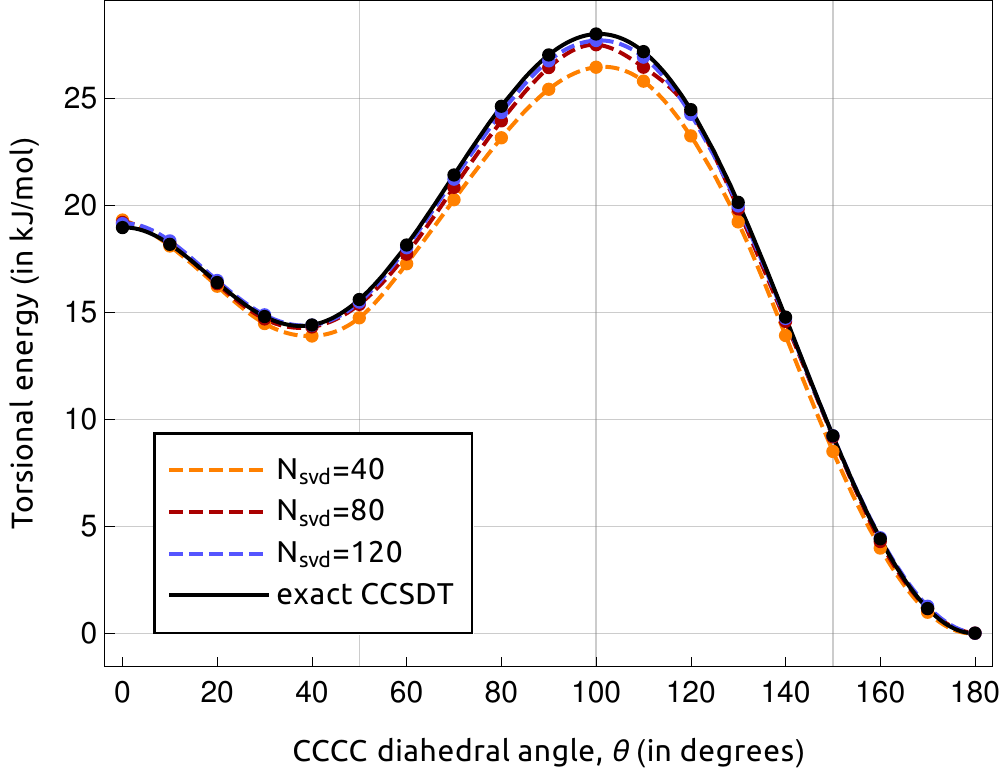} \\
 \includegraphics[scale=1.00]{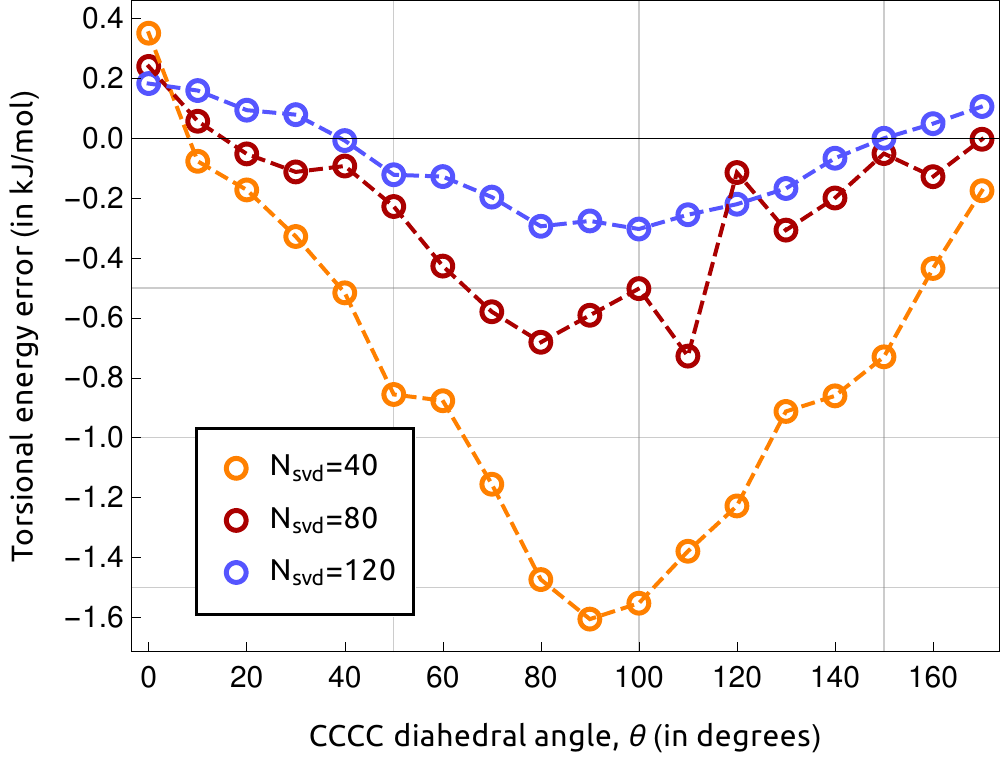} \\
\end{tabular}
 \caption{\label{buta} Torsional energy curve (upper panel) and torsional energy errors (lower panel) with respect to 
the exact CCSDT results calculated with the SVD-CCSDT method with $N_{\mathrm{SVD}}=40,80,120$ for butadiene molecule. 
The maximum size of the SVD subspace is 781.}
\end{figure}

We performed SVD-CCSDT calculations for the butadiene molecule (cc-pVDZ basis set) for the dihedral angles 
$\theta=0,10,20,\ldots,180^\circ$ with the rest of the geometry being the same as in the \emph{trans} conformer. In 
Table \ref{tab:buta1} we report mean absolute deviation and maximum deviation of the SVD-CCSDT torsional energies from 
the corresponding uncompressed CCSDT values for $\theta=0,10,20,\ldots,170^\circ$ and 
$N_{\mathrm{SVD}}=20,40,\ldots,120$. All results were arranged so that the \emph{trans} structure 
($\theta=180^\circ$) corresponds to zero energy level and is thus excluded from the error statistics. The results are 
also presented graphically in Fig. \ref{buta} where torsional energy curve and errors with respect to CCSDT are shown 
for each individual point. It is clear that even small SVD subspaces provide results that are close to the chemical 
accuracy. With $N_{\mathrm{SVD}}=60$ ($\rho\approx7.7\%$) all data points are already accurate to 1 kJ/mol or better, 
and $N_{\mathrm{SVD}}=120$ ($\rho\approx15.4\%$) gives results that, on average, are accurate to about 0.1 kJ/mol. The 
errors shown in Fig. \ref{buta} exhibit a regular and predictable behaviour without major jumps and discontinuities,
and there is only one minor exception occurring for $N_{\mathrm{SVD}}=80$ around $\theta=120^\circ$.

Comparison with analogous data obtained for the total energies leads to a conclusion that a systematic cancellation of 
errors occurs in the evaluation of relative SVD-CCSDT energies, thereby making the compressed coupled-cluster method 
even 
more beneficial under such circumstances. This is especially true for smaller SVD subspaces. Moreover, this example 
shows that the errors resulting from the truncation of the SVD subspace are weakly dependent on the geometry of the 
molecule provided that the same SVD subspace size is used consistently for all data points. This property is critical 
in computation of potentially surfaces that are smooth and regular, and thus can be fitted with a suitably chosen 
analytic functional form in order to, e.g., generate the molecular rotational/vibrational spectra or perform 
nuclear dynamics simulations.

\begin{table}[t]
  \caption{Parameters of the torsional energy curve (see text) for butadiene molecule determined with 
SVD-CCSDT and the uncompressed CCSDT methods (cc-pVDZ basis set). The energies are given in kJ/mol and angles in 
degrees.}
  \label{tab:buta2}
  \begin{tabular}{lrrrr}
    \hline
    parameter & \multicolumn{3}{c}{$N_{\mathrm{SVD}}$} & exact CCSDT \\
     & \multicolumn{1}{c}{40} & \multicolumn{1}{c}{80} & \multicolumn{1}{c}{120} &   \\
    \hline
    $\theta_{\rm gauche}$         & 39.4  & 37.6 & 37.6  & 37.6  \\
    $\theta_{\rm max}$           & 101.5 & 99.6 & 100.7 & 100.6 \\
    $\Delta E_{\rm barier}$      & 26.5  & 27.5 & 27.8  & 28.0  \\
    $\Delta E_{\rm gauche/trans}$ & 13.9  & 14.3 & 14.4  & 14.4  \\
    $\Delta E_{\rm cis/trans}$   & 19.3  & 19.2 & 19.1  & 19.0  \\
    \hline
  \end{tabular}
\end{table}

Lastly we assess the accuracy of some parameters that characterize the calculated torsional energy curve. These are:
\begin{itemize}
 \item values of the dihedral angle corresponding to the gauche structure ($\theta_{\rm gauche}$) and the maximum of 
the 
barrier ($\theta_{\rm max}$);
 \item the height of the barrier with respect to the trans structure ($\Delta E_{\rm barier}$);
 \item the energy difference between the gauche and trans structures ($\Delta E_{\rm gauche/trans}$) and between the 
cis 
and trans structures ($\Delta E_{\rm cis/trans}$).
\end{itemize}
For each method the above values were determined numerically with the help of $B$-splines interpolation of the 
calculated 
relative energies ($\theta=0,10,20,\ldots,180^\circ$). The results for $N_{\mathrm{SVD}}=40,80,120$ are shown in Table 
\ref{tab:buta2}. Already for $N_{\mathrm{SVD}}=120$ the difference between SVD-CCSDT and the exact CCSDT is probably 
smaller than the intrinsic accuracy of the latter. The obtained results agree reasonably well with the data available 
in the literature~\cite{engeln92,murcko96,sancho01,karpfen04,feller09b}.

\section{Conclusions and outlook}
\label{sec:conclusion}

We have reported an implementation of the full CCSDT electronic structure method using tensor 
decompositions to electron repulsion integrals and triple excitation amplitudes. The standard density-fitting 
approximation is used for the integrals while the triple amplitude tensor is represented in a Tucker-3 format. The 
quantities $U_{ai}^X$ used for the expansion in Eq. (\ref{tuck1}) are obtained by performing SVD of an approximate 
$t_{ijk}^{abc}$ tensor and retaining only those singular vectors that correspond to the largest 
singular values, as detailed previously~\cite{lesiuk19}. The compressed tensor $t_{XYZ}$ in Eq. 
(\ref{tuck1}) is obtained by performing coupled-cluster iterations within a subspace of triple excitations spanned by 
the chosen SVD basis.

The efficiency of this method relies on an observation that the optimal size of the SVD basis, $N_{\mathrm{SVD}}$, that 
is sufficient to 
deliver a constant relative accuracy of the correlation energy grows only linearly with the size of the system. This 
strategy allows to reduce the computational effort significantly, and leads to an approximate CCSDT method with 
practically $N^6$ 
scaling of its costs with the system size. This fact has been demonstrated by performing SVD-CCSDT calculations for 
linear alkanes with increasing chain length and analysing the computational timings.

The accuracy of the proposed method has been assessed by comparison with the exact (uncompressed) CCSDT. In the 
case of total energies it has been shown that for several small molecular systems the compression rates $\rho<20\%$ are 
more than sufficient to get chemically-accurate results. Taking into consideration that the compression rate $\rho$ is 
(asymptotically) inversely proportional to the size of the system, these results are very promising. In the case of 
relative energies we have observed a significant systematic cancellation of errors making accuracy levels below 1 
kJ/mol achievable with a considerably smaller $N_{\mathrm{SVD}}$ than for the total energies. The obtained relative 
energies exhibit a regular and predictable behaviour, without major jumps or discontinuities. We have also found that 
at present the practical accuracy limit of the SVD-CCSDT method is around $0.1$ kJ/mol due to oscillations caused by 
increasing numerical instabilities in the procedure of determining larger SVD subspaces. It is also worth pointing out 
that the SVD-CCSDT method preserves the appealing black-box nature of single-reference coupled-cluster theories and 
requires only one additional parameter ($N_{\mathrm{SVD}}$) to be specified by the user.

This work opens up a window for new developments in the field of ``compressed'' coupled-cluster theory. The most 
obvious extensions are perturbative methods that account for triple excitations outside the SVD subspace, as well as  
for quadruple excitations. Both types of corrections can be derived starting with the biorthogonal representation 
of the SVD-CCSDT state as the zeroth-order wavefunction. This formalism, related to the method-of-moments 
coupled-cluster theory~\cite{jankowski91}, was introduced by 
Stanton~\cite{stanton97} to explain the success of CCSD(T) over other methods that account for triple excitation 
perturbatively. Subsequently, a similar reasoning has been employed to derive systematic perturbative corrections for 
higher excitations~\cite{bomble05,kallay05,kallay08,eriksen14}.

Another interesting idea is solve for the so-called $\Lambda$ amplitudes~\cite{fritz86,salter89,jorgensen88,helga89} 
and evaluate the first-order properties such as multipole moments, nuclear gradients or electronic densities, from the 
coupled-cluster functional. Since in the CCSDT method $\Lambda$ and $T^\dagger$ are identical in the leading-order of 
perturbation theory, one can expect the SVD subspace used for $T$ to be adequate also for the expansion of $\Lambda$.
Application of the present tensor decomposition formalism to excited-state wavefunctions \emph{via} the 
equation-of-motion theory~\cite{sekino84,geertsen89,comeau93,stanton93} is also interesting due to capability of 
treating, e.g., doubly excited states. However, this is more complicated than computation of properties since 
determination of a proper SVD subspace requires to target one state at a time.

\begin{acknowledgement}
I would like to thank Dr. A. Tucholska and Prof. B. Jeziorski for fruitful discussions, and for reading and 
commenting on the manuscript. This work was supported by the National Science Center, Poland within 
the project 2017/27/B/ST4/02739.
\end{acknowledgement}

\begin{suppinfo}
The following files are available free of charge:
\begin{itemize}
  \item {\tt supp.pdf}: geometries of all molecules used as test cases in this work (XYZ or ZMAT formats).
\end{itemize}

\end{suppinfo}

\bibliography{svd_fullt}

\end{document}